\documentclass[aps,floatfix,showpacs,twocolumn,superscriptaddress,nofootinbib]{revtex4} 
 
\usepackage{graphicx} 
\usepackage{dcolumn} 
\usepackage{amsmath} 
 
\newcommand{\half}{{\textstyle{\frac{1}{2}}}}
\newcommand{\quarter}{{\textstyle{\frac{1}{4}}}}
\renewcommand{\case}[2]{\mbox{$\frac{#1}{#2}$}}

\begin{document} 
 
\title{%
Vector meson form factors and their quark-mass dependence}
 
\author{M.\ S.\ Bhagwat}
\affiliation{Physics Division, Argonne National Laboratory,
Argonne, IL 60439, U.S.A.}

\author{P.\ Maris}
\affiliation{Department of Physics and Astronomy, 
University of Pittsburgh, Pittsburgh, PA 15260, U.S.A.} 
                                       
\begin{abstract} 
The electromagnetic form factors of vector mesons are calculated in an
explicitly Poincar\'e covariant formulation, based on the
Dyson--Schwinger equations of QCD, that respects electromagnetic
current conservation, and unambiguously incorporates effects from
vector meson poles in the quark-photon vertex.  This method
incorporates a 2-parameter effective interaction, where the parameters
are constrained by the experimental values of chiral condensate and
$f_{\pi}$.  This approach has successfully described a large amount of
light-quark meson experimental data, e.g. ground state pseudoscalar
masses and their electromagnetic form factors; ground state vector
meson masses and strong and electroweak decays.  Here we apply it to
predict the electromagnetic properties of vector mesons.  The results
for the static properties of the $\rho$-meson are: charge radius
$\langle r_\rho^2 \rangle = 0.54~{\rm fm}^2$, magnetic moment $\mu =
2.01$, and quadrupole moment ${\cal Q} = -0.41$.  We investigate the
quark mass dependence of these static properties and find that our
results at the charm quark mass are in agreement with recent lattice
simulations.  The charge radius decreases with increasing quark mass,
but the magnetic moment is almost independent of the quark mass.
\end{abstract} 
 
\maketitle 

\section{Introduction}
Hadron form factors provide an important tool for understanding the
structure of bound states in QCD.  The coupling of a (virtual) photon
to a composite particle depends on its internal structure.  Even
static properties such as the charge radius and magnetic moment are
sensitive to the underlying QCD dynamics.  Thus it is not surprising
that there have been numerous studies of the electromagnetic form
factors of the nucleon, both theoretically and experimentally.  Also
the form factors of pseudoscalar mesons, in particular of the pion,
have been studied
extensively~\cite{Farrar:1979aw,Nesterenko:1982gc,Roberts:1994hh,Burden:1995ve,Choi:1997iq,Maris:1999bh,Maris:2000sk,Hwang:2001hj,Bakulev:2004cu,Volmer:2000ek,Fpinewexp}.
There are fewer studies of the vector meson form
factors~\cite{Cardarelli:1995yq,deMelo:1997hh,Hecht:1997uj,Hawes:1998bz},
though in recent years there has been a renewed interest in these form
factors~\cite{Samsonov:2003hs,Choi:2004ww,Braguta:2004kx,Aliev:2004uj,Dudek:2006ej,Lasscock:2006nh}.
However, the results of these theoretical studies appear to suffer
from a rather large model dependence, e.g. the results for the
quadrupole moment $G_{\rm Q}(0)$ differ by a factor of two between
different theoretical calculations, and the situation gets worse as
one moves away from $Q^2 = 0$~\cite{Aliev:2004uj}.

Here, we calculate the electromagnetic form factors of the $\rho$ and
both the neutral and charged $K^{\star}$ mesons using the model
proposed in Ref.~\cite{Maris:1999nt}.  The parameters of this model
were adjusted to reproduce the experimental values for the chiral
condensate, the pion mass and decay constant, and the kaon mass;
calculations of other mesonic observables are predictions.  The
obtained vector meson masses and decay constants are in agreement with
data~\cite{Maris:1999nt} as are the results for the strong decays of
the vector mesons~\cite{Jarecke:2002xd}.  The pion and kaon
electromagnetic form factors~\cite{Maris:1999bh,Maris:2000sk} are also
in excellent agreement with
experiments~\cite{Volmer:2000ek,Fpinewexp}, as are the transition form
factors such as the $\rho\to\pi \gamma$~\cite{Maris:2002mz}.  This
proclivity of theoretical results to experimental data provides good
reason to expect that the model will accurately describe the electromagnetic
form factors of light vector mesons.

In Sec.~\ref{Sec:DSE} we review the Dyson--Schwinger equations used to
calculate the quark propagators and the meson Bethe-Salpeter
amplitudes.  Next, we discuss the truncation and the model for the
effective interaction, and the results for the meson masses, decay
constants, and their quark mass dependence.  We explicitly demonstrate
frame independence: our results for physical observables are
independent of the total meson momentum.  This is important since in
form factor calculations at least one of the mesons is moving.  In
Sec.~\ref{Sec:EMff}, we briefly discuss the general form of vector
meson form factors and the method of calculation; an essential element
of our calculation is the treatment of the quark-photon vertex.  We
present our numerical results for the form factors of the $\rho$ and
of both the neutral and charged $K^\star$ mesons in
Sec.~\ref{Sec:Num}, as well as the mass dependence of the static
electromagnetic properties of equal-mass vector mesons; we also
compare our results with other calculations and with recent lattice
simulations.  Finally, some of the details of our calculation are
given in the appendix.

\section{Dyson--Schwinger equations of QCD
\label{Sec:DSE}}
The Dyson--Schwinger equations [DSEs] are the equations of motion of a
quantum field theory.  They form an infinite hierarchy of coupled
integral equations for the Green functions ($n$-point functions) of
the theory.  Bound states (mesons, baryons) appear as poles in the
Green functions.  Thus, a study of the poles in $n$-point functions
using the set of DSEs will tell us about properties of hadrons.  For
recent reviews on the DSEs and their use in hadron physics, see
Refs.~\cite{Roberts:2000aa,Alkofer:2000wg,Maris:2003vk,Fischer:2006ub}.

\subsection{Quark propagator}
The exact DSE for the quark propagator is\footnote{We use
Euclidean metric $\{\gamma_\mu,\gamma_\nu\} = 2\delta_{\mu\nu}$,
$\gamma_\mu^\dagger = \gamma_\mu$ and $a\cdot b = \sum_{i=1}^4 a_i
b_i$.}
\begin{eqnarray}
S(p)^{-1}  &=&  i \not\!p\, Z_2+ m_q(\zeta)\,Z_4 + {}
\nonumber \\ && Z_1 \int_k \, g^2D_{\mu \nu}(q)
        \, \gamma_\mu {\hbox{$\frac{\lambda^i}{2}$}}
        \, S(k) \, \Gamma^i_\nu(k,p)  \,,
\label{Eq:quarkDSE}
\end{eqnarray}
where $D_{\mu\nu}(q=k-p)$ is the renormalized dressed gluon
propagator, and $\Gamma^i_\nu(k,p)$ is the renormalized dressed
quark-gluon vertex.  The notation $\int_k$ stands for $\int^\Lambda
d^4 k/(2\pi)^4$.  For divergent integrals a translationally invariant
regularization is necessary; the regularization scale $\Lambda$ is to
be removed at the end of all calculations, after renormalization, and
will be suppressed henceforth.

The solution of Eq.~(\ref{Eq:quarkDSE}) can be written as
\begin{eqnarray}
 S(p)  &=&  \frac{Z(p^2)}{i \not\!p + M(p^2)} \,,
\label{Eq:quarkprop}
\end{eqnarray}
renormalized according to $S(p)^{-1} = i\,/\!\!\!p + m(\zeta)$ at a
sufficiently large spacelike $\zeta^2$, with $m(\zeta)$ the current quark
mass at the scale $\zeta$.  Both the propagator, $S(p)$, and the vertex,
$\Gamma^i_\mu$, depend on the quark flavor, although we have not
indicated this explicitly.  The renormalization constants $Z_2$ and
$Z_4$ depend on the renormalization point and on the regularization
mass-scale, but not on flavor: in our analysis we employ a
flavor-independent renormalization scheme.

\subsection{Mesons}
Bound states correspond to poles in $n$-point functions: for example a
meson appears as a pole in the 2-quark, 2-antiquark Green function
$G^{(4)}=\langle 0 | q_1 q_2 \bar{q_1} \bar{q_2} | 0 \rangle$.  In the
vicinity of a meson, i.e. in the neighborhood of $P^2 = -M^2$ with $M$
being the meson mass, such a Green function behaves like
\begin{eqnarray}
 G^{(4)} &\sim& 
 \frac{\chi(p_{\hbox{\scriptsize out}}, p_{\hbox{\scriptsize in}};P) \; 
  \bar{\chi}(k_{\hbox{\scriptsize in}},k_{\hbox{\scriptsize out}};P)}{P^2 + M^2} \,,
\end{eqnarray}
where $P$ is the total 4-momentum of the meson, $p_{\hbox{\scriptsize
out}}$ and $p_{\hbox{\scriptsize in}}$ are the 4-momenta of the
outgoing quark and incoming quark respectively, and similarly for
$k_{\hbox{\scriptsize in}}$ and $k_{\hbox{\scriptsize out}}$.
Momentum conservation relates these momenta: $p_{\hbox{\scriptsize
out}}-p_{\hbox{\scriptsize in}} = P = k_{\hbox{\scriptsize out}} -
k_{\hbox{\scriptsize in}}$.

The function $\chi(p_{\hbox{\scriptsize out}},p_{\hbox{\scriptsize
in}};P)$ describes the coupling of the bound state to a dressed quark
and antiquark.  It satisfies the homogeneous Bethe--Salpeter equation
[BSE]
\begin{eqnarray}
\lefteqn{ \Gamma(p_{\hbox{\scriptsize out}}, p_{\hbox{\scriptsize in}};P) \; = }
\nonumber \\ &&  
 \int_k \! K(p_{\hbox{\scriptsize out}},p_{\hbox{\scriptsize in}};
             k_{\hbox{\scriptsize out}},k_{\hbox{\scriptsize in}}) \,
	\chi(k_{\hbox{\scriptsize out}}, k_{\hbox{\scriptsize in}};P) \, ,
\label{Eq:genBSE}
\end{eqnarray}
at discrete values $P^2 = -M^2$ of the total meson 4-momentum $P$.
Here $\Gamma$ is the Bethe--Salpeter amplitude [BSA]
\begin{eqnarray}
\Gamma(k_{\hbox{\scriptsize out}},k_{\hbox{\scriptsize in}}; P) \, = \,
   S(k_{\hbox{\scriptsize out}})^{-1}  
   \chi(k_{\hbox{\scriptsize out}},k_{\hbox{\scriptsize in}}; P)  
   S(k_{\hbox{\scriptsize in}})^{-1} \,,
\end{eqnarray}
and the kernel $K$ is the $q \bar{q}$ irreducible quark-antiquark
scattering kernel.

The meson BSA is normalized according to
\begin{eqnarray}
 \lefteqn{2\, P_\mu = N_c \; \frac{\partial}{\partial Q_\mu} \bigg\{ }
\nonumber\\ && \int_{k,q} {\rm Tr}\big[
          \bar\chi(k_{\hbox{\scriptsize in}},k_{\hbox{\scriptsize out}})\,
	  K(\tilde{k}_{\hbox{\scriptsize out}},\tilde{k}_{\hbox{\scriptsize in}};
	  \tilde{q}_{\hbox{\scriptsize out}},\tilde{q}_{\hbox{\scriptsize in}})\,
          \chi(q_{\hbox{\scriptsize out}},q_{\hbox{\scriptsize in}}) \big] 
\nonumber\\ && {} + 
\int_k {\rm Tr}\big[
        \bar\Gamma(k_{\hbox{\scriptsize in}},k_{\hbox{\scriptsize out}})\,
	  S(\tilde{k}_{\hbox{\scriptsize out}})\,
            \Gamma(k_{\hbox{\scriptsize out}},k_{\hbox{\scriptsize in}})\,
	    S(\tilde{k}_{\hbox{\scriptsize in}}) \big]
\bigg\} \bigg|_{Q=P}
\nonumber \\
\label{Eq:BSEnorm}
\end{eqnarray}
at $P^2=-M^2$, with $\tilde{k}_{\hbox{\scriptsize
out}}-\tilde{k}_{\hbox{\scriptsize in}}=Q=
\tilde{q}_{\hbox{\scriptsize out}}-\tilde{q}_{\hbox{\scriptsize in}}$.
The properly normalized BSA $\Gamma(p_{\hbox{\scriptsize
out}},p_{\hbox{\scriptsize in}};P)$ [or equivalently,
$\chi(p_{\hbox{\scriptsize out}},p_{\hbox{\scriptsize in}};P)$]
completely describes the meson as a $q\bar{q}$ bound state.  Mesons of
different spins and parity are characterized by different Dirac
structures, e.g. the BSA of massive vector mesons can be decomposed
into eight Dirac structures~\cite{Maris:1999nt}
\begin{eqnarray}
\lefteqn{ \Gamma^{\rm V}_\mu(k+\eta P,k-(1-\eta)P; P) = }
\nonumber \\
        && \sum_{i=1}^8 f^i(k^2, k\cdot P; \eta) \, T^i_\mu(k,P) \,,
\label{Eq:decompV}
\end{eqnarray}
where $T^i_\mu(k,P)$ are eight independent transverse Dirac tensors.
The invariant amplitudes $f^i$ are Lorentz scalar functions of $k^2$
and $k\cdot P$, and depend on the momentum partitioning parameter
$\eta$.  Physical observables however are independent of $\eta$.

\subsection{Rainbow-ladder truncation}
A viable truncation of the infinite set of DSEs has to respect
relevant (global) symmetries of QCD such as chiral symmetry,
Poincar\'e covariance, and renormalization group invariance.  For
electromagnetic interactions the truncation should also respect
current conservation.  These properties are built into the
rainbow-ladder
truncation~\cite{Maris:1999bh,Maris:2000sk,systematicexp,Maris:1997hd,Maris:1997tm}.
In this scheme, the kernel $K$ of the meson BSE is replaced by an
(effective) one-gluon exchange
\begin{eqnarray}
\lefteqn{ K(p_{\hbox{\scriptsize out}},p_{\hbox{\scriptsize in}};
   k_{\hbox{\scriptsize out}},k_{\hbox{\scriptsize in}}) }
\nonumber \\ &\to &
        -4\pi\,\alpha(q^2)\, D_{\mu\nu}^{\rm free}(q)
        \textstyle{\frac{\lambda^i}{2}}\gamma_\mu \otimes
        \textstyle{\frac{\lambda^i}{2}}\gamma_\nu \,,
\end{eqnarray}
where $q=p_{\hbox{\scriptsize out}}-k_{\hbox{\scriptsize
out}}=p_{\hbox{\scriptsize in}}-k_{\hbox{\scriptsize in}}$, and
$\alpha(q^2)$ is an effective running coupling; $D_{\mu\nu}^{\rm
free}(q)$ is the free gluon propagator; and we choose to work in
Landau gauge.  The corresponding rainbow truncation of the quark DSE
is
\begin{eqnarray}
Z_1 g^2 D_{\mu \nu}(q) \Gamma^i_\nu(k,p) &\to &
 4\pi\,\alpha(q^2) \, D_{\mu\nu}^{\rm free}(q)\, \gamma_\nu
                                        \textstyle\frac{\lambda^i}{2} \,.
\label{Eq:rainbow}
\end{eqnarray}
This truncation is the first term in a systematic
expansion~\cite{systematicexp} of the quark-antiquark scattering
kernel $K$; asymptotically, it reduces to leading-order perturbation
theory.  Furthermore, these two truncations are mutually consistent in
the sense that the combination produces vector and axial-vector
vertices satisfying their respective Ward identities.  In the axial
case, this ensures that in the chiral limit the ground state
pseudoscalar mesons are the massless Goldstone bosons associated with
chiral symmetry breaking~\cite{Maris:1997hd,Maris:1997tm}.  In the
vector case, this ensures, in combination with impulse approximation,
electromagnetic current conservation~\cite{Maris:1999bh,Maris:2000sk}.

\section{Model calculations
\label{Sec:model}}
The ultraviolet behavior of the effective running coupling is dictated
by the one-loop renormalization group equation; the infrared behavior
of the effective interaction is modeled, and constrained by
phenomenology.  Here, we employ the model of Ref.~\cite{Maris:1999nt}
for $\alpha(q^2)$
\begin{eqnarray}
\label{gvk2}
\frac{{4\pi\alpha}(q^2)}{k^2} &=&
        \frac{4\pi^2\, D \,k^2}{\omega^6} \, {\rm e}^{-k^2/\omega^2}
\nonumber \\ && {}
        + \frac{ 4\pi^2\, \gamma_m \; {\cal F}(k^2)}
        {\textstyle{\frac{1}{2}} \ln\left[\tau + 
        \left(1 + k^2/\Lambda_{\rm QCD}^2\right)^2\right]} \;,
\end{eqnarray}
with \mbox{${\cal F}(s)=(1 - \exp\frac{-s}{4 m_t^2})/s$},
\mbox{$\gamma_m=12/(33-2N_f)$}, and fixed parameters
\mbox{$m_t=0.5\,{\rm GeV}$}, \mbox{$\tau={\rm e}^2-1$},
\mbox{$N_f=4$}, and \mbox{$\Lambda_{\rm QCD} = 0.234\,{\rm GeV}$}.
The remaining parameters $\omega$ and $D$ were fitted in
Ref.~\cite{Maris:1999nt} to reproduce $f_\pi$ and the chiral
condensate: $\omega = 0.4~{\rm GeV}$ and $D = 0.93~{\rm GeV}^2$.  

\subsection{Results for light quarks}
With this model, we obtain good agreement with the experimental values
for the light pseudoscalar and vector meson masses and leptonic decay
constants, see Table~\ref{Table:model}.  The current quark masses
$m_{u/d} = 3.7~{\rm MeV}$ and $m_s = 83.8~{\rm MeV}$ at the
renormalization point $\zeta = 19~{\rm GeV}$ were
fitted~\cite{Maris:1999nt} to the pion and kaon mass respectively.
Using the one-loop expression to evolve these masses down to $\zeta =
2~{\rm GeV}$ gives $m_{u/d}(1~{\rm GeV})= 5.0~{\rm MeV}$ and
$m_s(2~{\rm GeV}) = 118~{\rm MeV}$.
\begin{table}[b]
\caption{DSE results~\cite{Maris:1999nt,Maris:2006ea} for the
pseudoscalar and vector meson masses and decay constants, together
with experimental data from Ref.~\cite{PDG}, unless indicated
otherwise.  All entries are in GeV.
\label{Table:model} }
\renewcommand{\tabcolsep}{1pc} 
\renewcommand{\arraystretch}{1.2} 
\begin{tabular}{@{}lll}
\hline
        & experiment   & calculated        \\
        & (estimates)  & ($^\dagger$ fitted) \\ \hline
$m_{u/d}(\zeta=2~{\rm GeV})$ & 0.003 to 0.006 & 0.005     \\
$m_{s}(\zeta=2~{\rm GeV})$ & 0.095(25)  & 0.118    \\ \hline
$m_{c}(\zeta=m_c) $ & 1.25(9)  & 1.30     \\ \hline
$M_\pi$         &  0.135, 0.140  &    0.138$^\dagger$ \\
$f_\pi$         &  0.131         &    0.131$^\dagger$  \\
$M_K$           &  0.496         &   0.497$^\dagger$ \\
$f_K$           &  0.160         &   0.155        \\ \hline
$M_\rho$, $M_\omega$        
                &  0.776, 0.783  &   0.742        \\
$f_\rho$, $f_\omega$        
                &  0.221(2), 0.195(4) &   0.207        \\
$M_{K^\star}$   &  0.892         &   0.936        \\
$f_{K^\star}$   &  0.224(11)     &   0.241        \\
$M_\phi$        &  1.020         &   1.074        \\
$f_\phi$        &  0.229(4)      &   0.259        \\ \hline
$M_{\eta_c}$    &  2.980         &   2.91         \\
$f_{\eta_c}$    &  0.335(75)~\cite{Edwards:2000bb}  & 0.38 \\
$M_{J/\Psi}$    &  3.097         &   3.10$^\dagger$ \\
$f_{J/\Psi}$    &  0.416(6)      &   0.42   \\ \hline
\end{tabular}\\[2pt]
\end{table}

These results show little sensitivity to variations in the model
parameters~\cite{refCDR}, as long as the integrated strength of the
effective interaction is strong enough to generate an acceptable
amount of chiral symmetry breaking, as indicated by the chiral
condensate.  This is not true for heavier states consisting of light
quarks: e.g. the radially excited pion is quite sensitive to details
of the interaction~\cite{Holl:2005vu}.

Not only the meson masses and leptonic decay constants, but also a
wide range of other observables agree with experiments, without
adjusting any of the parameters, see~\cite{Maris:2003vk} and
references therein.  In particular
$F_\pi(Q^2)$~\cite{Maris:1999bh,Maris:2000sk,Volmer:2000ek,Fpinewexp}
and the $\rho$-$\pi$-$\gamma$, $\omega$-$\pi$-$\gamma$, and
$\pi$-$\gamma\gamma$ form factors~\cite{Maris:2002mz} are well
described by this model.  We therefore expect the model to describe
the electromagnetic form factors of light vector mesons quite
accurately as well.

\begin{figure}[tb]
\includegraphics[width=0.96\columnwidth]{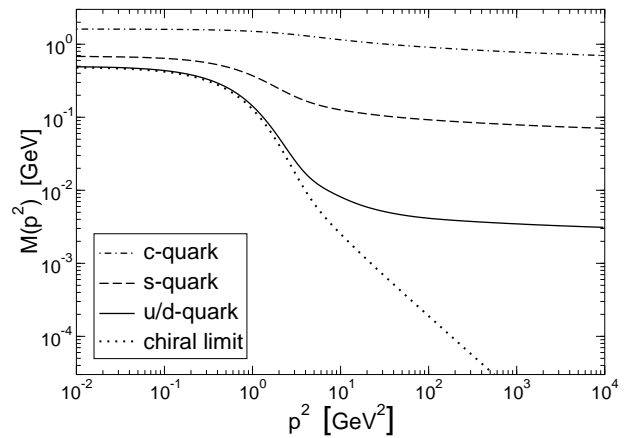}
\caption{Dynamical quark mass function using the rainbow-ladder
truncation of Ref.~\cite{Maris:1999nt}.
\label{Fig:quarkprp}}
\end{figure}
The corresponding quark propagator functions are shown in
Fig.~\ref{Fig:quarkprp}.  These predictions for the quark mass
function have been semi-quantitatively confirmed in recent lattice
simulations of
QCD~\cite{Skullerud:2001aw,Bowman:2002kn,Bowman:2004xi}.  Point-wise
agreement for a range of quark masses requires this interaction to be
flavor-dependent~\cite{Bhagwat:2003vw}, and dressing the quark-gluon
vertex $\Gamma^i_\nu(q,p)$ ensures this dependence.  The consequences
of a dressed vertex for the meson BSEs are also currently being
explored and indications are that in the pseudoscalar and vector
channels, the effects are
small~\cite{systematicexp,Bhagwat:2004hn,Matevosyan:2006bk}.  The
non-trivial infrared structure of the quark-gluon vertex is under
investigation using both lattice
simulations~\cite{Skullerud:2002ge,Skullerud:2003qu,Skullerud:2004gp}
and non-perturbative DSE
methods~\cite{Bhagwat:2004kj,Llanes-Estrada:2004jz}, but these studies
are not yet conclusive.

\subsection{Quark-mass dependence}
\begin{figure}[tb]
\includegraphics[width=0.96\columnwidth]{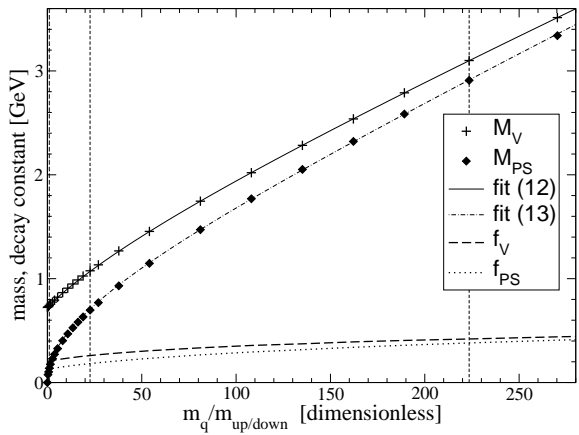}
\caption{ 
Pseudoscalar and vector meson masses and leptonic decay constants as
function of the current quark mass (normalized to the physical up/down
quark mass).  Vertical dashed lines indicate the up/down, strange, and
charm quark masses.
\label{Fig:meson}}
\end{figure}

In recent years, this model has been
extended~\cite{Maris:2006ea,Maris:2005tt} to the charm and bottom quarks as
well, in an attempt to describe both the light mesons and heavy
quarkonia within one framework.  In Fig.~\ref{Fig:meson} we show our
results for the vector meson mass and decay constant as function of
the current quark mass; for comparison we also include the evolution
of the corresponding pseudoscalar meson mass and decay constant.  In
this and subsequent figures, we normalize the current quark masses by
the physical up and down quark masses of our model; for the strange
quark we have $m_s \approx 23~m_{u/d}$.

The charm quark mass $m_c = 0.827~{\rm GeV}$ at the renormalization
point $\zeta = 19~{\rm GeV}$ is fixed by the experimental value of
$J/\Psi$ mass.  Again, using one-loop evolution this corresponds to
$m_c = 1.30~{\rm GeV}$ at $\zeta = m_c$, see Table~\ref{Table:model}.
The decay constant for the $J/\Psi$ is in good agreement with data,
and the resulting post-dictions for the $\eta_c$ mass and decay
constant are in reasonable agreement with the data as well.

On a limited domain, both the vector and the pseudoscalar meson masses
can be fitted reasonably well by
\begin{eqnarray}
  M_{\rm meson}^2 &=& C_0 + C_1 \; m_q + C_2 \; m_q^2 \;,
\end{eqnarray}
where $m_q$ is the current quark mass at our renormalization point
$\zeta = 19~{\rm GeV}$.  For the mass region we are interested in here,
from the chiral limit up to the charm quark mass, the fit parameters are
\begin{eqnarray}
M_{\rm PS}: &&  C_0 = 0, \quad C_1 = 5.49, \quad C_2 = 5.77, 
\\
M_{\rm V}: &&  C_0 = 0.53, \quad C_1 = 6.93, \quad C_2 = 4.88 \,,
\end{eqnarray}
as illustrated in Fig.~\ref{Fig:meson}.  For larger quark masses, the
fit parameters for the pseudoscalar and vector mesons become closer
and closer to each other: a fit on the domain $m_c < m_q < 2\,m_b$
gives $C_0=-1.3$ and $C_1=7.92$ for the pseudoscalar mesons, and
$C_0=-0.2$ and $C_1=8.02$ for the vector mesons, with a common
parameter $C_2 = 4.46$.  This reflects the fact that in the
heavy-quark limit, the pseudoscalar and vector mesons become
degenerate: in the limit $m_q \to \infty$ the above fit gives
$M_{\rm V} - M_{\rm PS} \to \half (C_1^{\rm V}-C_1^{\rm PS})/\sqrt{C_2} \approx 0$.

The leptonic decay constants increase with the current quark mass,
both for the pseudoscalar and for the vector mesons.  Based on the
experimental partial decay width of vector mesons, it was
conjectured~\cite{Yennie:1974ga} that the vector meson decay constants
increase with quark mass as $f_{\rm V} \propto \sqrt{M_{\rm V}}$.  On
the other hand~\cite{Achasov:2001sn}, Coulomb-potential models
typically give $f_{\rm V} \propto M_{\rm V}$, whereas a linear
(confining) potential produces $f_{\rm V}\sim$~constant.  Our
numerical results suggest that both $f_{\rm V}$ and $f_{\rm PS}$
increase approximately linearly with quark mass, at least for masses
in the $m_c$ to twice $m_b$ range~\cite{Bhagwat:2006xi}.  This is
Coulomb-potential-like behavior, which may be natural since the
effective interaction, Eq.~\ref{gvk2}, reduces to one-gluon exchange
in the ultraviolet region.  However, further investigations are needed
in order to determine the true asymptotic behavior of the decay
constants.

\subsection{Frame independence}
The BSE is usually solved in the rest-frame of the meson.  However,
the calculation of electromagnetic form factors in any reference frame
entails either the initial meson, or the final meson, or both, to have
non-zero 3-momentum.  In a method that is not Poincar\'e covariant the
wave functions for the moving meson would have to be boosted.  One of
the advantages of the DSE approach to hadron physics is its manifest
Poincar\'e covariance.

\begin{figure}[tb]
\includegraphics[width=0.92\columnwidth]{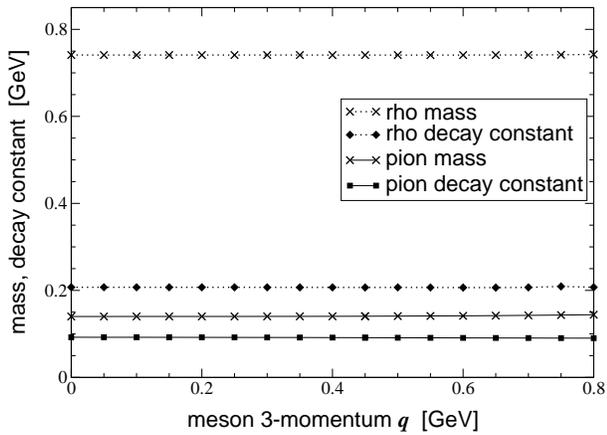}
\caption{Pion and $\rho$ mass and decay constant calculated in a
moving frame, as function of the meson 3-momentum (Figure adapted
from Ref.~\cite{Maris:2005tt}).
\label{Fig:framedep}}
\end{figure}

As an explicit demonstration, we calculate the static $\pi$ and $\rho$
properties in a moving frame\footnote{In the Euclidean metric that we
are using here, the rest-frame is characterized by $P_\mu=(0,0,0,i\,M)$.} 
$P_\mu = (q,0,0,i\,E)$ where $q$ is the 3-momentum of the
moving meson~\cite{Maris:2005tt}.  Within this frame we solve again
the homogeneous BSE, Eq.~(\ref{Eq:genBSE}), and calculate the
corresponding electroweak decay constant.  Numerically this is 
a demanding task, since the Lorentz scalar functions of
Eq.~(\ref{Eq:decompV}) are now functions of a radial variable $k^2$
and {\em two} angles
\begin{eqnarray}
 k\cdot P &=& i\, k\,E \cos\alpha + k\,q\, \sin\alpha \cos\beta \,,
\end{eqnarray}
and the integral equation has to be solved in the three independent
variables $k^2$, $\alpha$, and $\beta$.  With current computer
resources, this can be done without further approximations, and the
results, shown in Fig.~\ref{Fig:framedep}, are indeed independent of
the meson 3-momentum, illustrating that this approach is indeed
Poincar\'e covariant.  We can now use this same approach to calculate
meson form factors in an explicitly covariant manner.

\section{Electromagnetic form factors
\label{Sec:EMff}}

\subsection{Vector meson form factors}
Consider the electromagnetic current of a vector meson with incoming
meson momentum $P^-_\rho=P_\rho - \half Q_\rho$, outgoing meson
momentum $P^+_\sigma=P_\sigma + \half Q_\sigma$, and incoming photon
momentum $Q_\mu$.  With this notation, the general form for the
coupling of a photon to a vector meson can be written
as~\cite{Hawes:1998bz,Arnold:1979cg}
\begin{eqnarray}
 \Lambda_{\mu, \rho\sigma}(P, Q) &=& 
  - \sum_{j=1}^{3} T_{\mu\,\rho\sigma}^{[j]}(P, Q) \; F_{j}(Q^2) \; ,
\\
 T_{\mu, \rho\sigma}^{[1]}(P, Q) &=&
 2 \, P_\mu \, {\cal P}_{\rho\gamma}^T(P^-)
               {\cal P}_{\gamma\sigma}^T(P^+) \;,
\label{Eq:HPT1}  \\
 T_{\mu, \rho\sigma}^{[2]}(P, Q) &=&
\left(Q_\rho - P^-_\rho\frac{Q^2}{2\;M^2}\right) 
   \; {\cal P}_{\mu\sigma}^T(P^+) 
\nonumber \\ && {} -
\left(Q_\sigma + P^+_\sigma\frac{Q^2}{2\;M^2}\right) 
   \; {\cal P}_{\mu\rho}^T(P^-) \;,
\label{Eq:HPT2}  \\
 T_{\mu, \rho\sigma}^{[3]}(P, Q) &=&
\frac{P_\mu}{M^2} \;
\left(Q_\rho - P^-_\rho\frac{Q^2}{2\;M^2}\right) \;
\nonumber \\ && {} \times
\left(Q_\sigma + P^+_\sigma\frac{Q^2}{2\;M^2}\right) \;, 
\label{Eq:HPT3}
\end{eqnarray}
where
\begin{eqnarray}
  {\cal P}_{\mu\nu}^T(k) &=& \delta_{\mu\nu} - \frac{k_\mu k_\nu}{k^2}
\end{eqnarray}
is the transverse projector.  The vector meson is on-shell: $(P^-)^2 =
(P^+)^2 = -M^2$, where $M$ is the mass of the vector meson, and thus
$P^2 + \quarter Q^2 = -M^2$ and $P\cdot Q = 0$.  This coupling obeys
the following relations
\begin{eqnarray}
 P^+_\rho \Lambda_{\mu\rho\sigma}(p, p') &=& 0  \; ,
\\
 P^-_\sigma \Lambda_{\mu\rho\sigma}(p, p') &=& 0 \; ,
\\
 Q_\mu \Lambda_{\mu\rho\sigma}(p, p') &=& 0 \; .
\end{eqnarray}
The first two equations simply reflect that the (massive) vector
mesons are transverse; the last equation follows from current
conservation.

The electric, magnetic, and quadrupole form factors $G_{\rm E}, G_{\rm
M}, G_{\rm Q}$ can be expressed in terms of these scalar functions $F_i$
\begin{eqnarray}
G_{\rm E}(Q^2) &=& F_{1}(Q^2) 
                + \frac{2}{3}\,\frac{Q^{2}}{4\, M^{2}} G_{\rm Q}(Q^2) ~,
\label{Eq:HPGE}  \\
G_{\rm M}(Q^2) &=& -F_{2}(Q^2) ~,
\label{Eq:HPGM}  \\
G_{\rm Q}(Q^2) &=& F_{1}(Q^2) + F_{2}(Q^2) 
  + \left( 1+\frac{Q^{2}}{4\,M^{2}} \right) F_{3}(Q^2) ~.
\nonumber \\
\label{Eq:HPGQ}
\end{eqnarray}
The electric monopole moment (i.e. the electric charge), magnetic
dipole moment and the electric quadrupole moment follow from the
values of these form factors in the limit $Q^{2}\to 0$ 
\begin{eqnarray}
G_{\rm E}(Q^{2}=0) &=& 1  \;,
\\
G_{\rm M}(Q^{2}=0) &=& \mu  \;,
\\
G_{\rm Q}(Q^{2}=0) &=& {\cal Q} \;.
\end{eqnarray}
Here the magnetic moment $\mu$ and the quadrupole moment $\cal Q$ are
introduced; the electric charge is 1 in terms of the fundamental
charge unit $e$.  For point-like vector particles, the magnetic and
quadrupole moments are $\mu = 2$ in units of $e/2M_{\rm V}$ and ${\cal
Q} = -1$ in units of $e/M^{2}_{\rm V}$,
respectively~\cite{Brodsky:1992px}.

\subsection{Impulse approximation}
\begin{figure}[tb]
\includegraphics[width=5cm,angle=90]{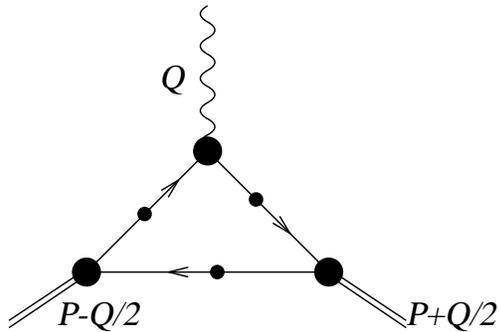}
\caption{Meson form factor in impulse approximation
\label{Fig:impulse}}
\end{figure}

The generalized impulse approximation allows electromagnetic processes
to be described in terms of dressed quark propagators, bound state
BSAs, and the dressed $q\bar{q}\gamma$-vertex, see
Fig.~\ref{Fig:impulse}.  In combination with ladder-rainbow truncation
for the vertices and the quark propagators, it ensures electromagnetic
current conservation~\cite{Maris:1999bh,Maris:2000sk}, see also
Sec.~\ref{subsec:vertex} below.  Phenomenologically, this
approximation has proved to be very successful in describing the pion
electromagnetic form factor~\cite{Volmer:2000ek,Fpinewexp}.

Consider for example the 3-point function describing the coupling of
a photon with momentum $Q$ to a vector meson $a\bar{b}$, with initial
and final momenta \mbox{$P \pm Q/2$}.  This interaction can be written
as the sum of two terms
\begin{eqnarray}
\Lambda^{a\bar{b}}_{\mu,\ \rho\sigma}(P,Q) &=&
                \hat{Q}^a \, \Lambda^{aa\bar{b}}_{\mu,\ \rho\sigma}
                + \hat{Q}^{\bar{b}} \, \Lambda^{a\bar{b}\bar{b}}_{\mu,\ \rho\sigma} \,,
\label{Eq:twoterms}
\end{eqnarray}
where $\hat{Q}$ is the quark or antiquark electric charge, and where
$\Lambda^{a\bar{b}a}(P,Q)$ and $\Lambda^{a\bar{b}\bar{b}}(P,Q)$
describe the coupling of a photon to the quark ($a$) and antiquark
($\bar{b}$) respectively.  In impulse approximation, these couplings
are given by
\begin{eqnarray}
\Lambda^{aa\bar{b}}_{\mu,\ \rho\sigma}(P,Q) &=&
        i\,N_c \int_k \!{\rm Tr}\big[ \Gamma^{a}_\mu(q_-,q_+)
	  \, \chi_\rho^{a\bar{b}}(q_+,q) 
\nonumber \\ && {} \times
        S^b(q)^{-1} \, \bar\chi_\sigma^{\bar{b}a}(q,q_-)\big] \;,
\label{Eq:triangle}
\end{eqnarray}
with \mbox{$q = k-P/2$} and \mbox{$q_\pm = k+P/2 \pm Q/2$}, 
and similarly for $\Lambda^{a\bar{b}\bar{b}}_{\mu,\ \rho\sigma}$.  

For the $\rho$ mesons it is sufficient to calculate the coupling of
the photon to a single quark, a direct consequence of isospin
invariance, while for the $K^\star$ mesons we add contributions from
photon coupling to the quark and the antiquark.  Thus for the charged
$K^\star$ the form factors are given by
\begin{eqnarray}
  F_i^{K^{\star,+}}(Q^2) &=& 
    \case{2}{3} F_i^{uu\bar{s}}(Q^2)
   - \case{1}{3} F_i^{u\bar{s}\bar{s}}(Q^2)
\\
   &=& \case{2}{3} F_i^{uu\bar{s}}(Q^2)
      + \case{1}{3} F_i^{ss\bar{u}}(Q^2)  \;,
\end{eqnarray}
and similarly for the neutral $K^\star$ we have
\begin{eqnarray}
  F_i^{K^{\star,0}}(Q^2) 
   &=& -\case{1}{3} F_i^{uu\bar{s}}(Q^2)
      + \case{1}{3} F_i^{ss\bar{u}}(Q^2) \;.
\end{eqnarray}

\subsection{Quark-photon vertex
\label{subsec:vertex}}
The impulse approximation and rainbow-ladder truncation 
together lead to current conservation
only if the quark-photon vertex $\Gamma_\mu$ satisfies the
vector Ward--Takahashi identity~\cite{Roberts:1994hh}
\begin{eqnarray}
 i\,Q_\mu \,\Gamma_{\mu}(k_+,k_-;Q) & = & S^{-1}(k+\half Q) - S^{-1}(k-\half Q) \;.
\nonumber \\ 
\end{eqnarray}
This identity can be satisfied by the Ball--Chiu
Ansatz~\cite{Ball:1980ay} for the quark-photon vertex; a more
consistent approach, which we follow, is to use the solution of the
inhomogeneous BSE for the $q\bar{q}\gamma$-vertex in the ladder
truncation~\cite{Maris:1999bh,Maris:2000sk}
\begin{eqnarray}
\lefteqn{\Gamma_\mu(p_{\hbox{\scriptsize out}},p_{\hbox{\scriptsize in}};Q) 
       \;=\; Z_2\, \gamma_\mu + \int_k  
     K(p_{\hbox{\scriptsize out}},p_{\hbox{\scriptsize in}};
       k_{\hbox{\scriptsize out}},k_{\hbox{\scriptsize in}})}
\nonumber \\ && {}\times
          S(k_{\hbox{\scriptsize out}})\,
 \Gamma_\mu(k_{\hbox{\scriptsize out}},k_{\hbox{\scriptsize in}};Q) \, 
          S(k_{\hbox{\scriptsize in}}) \;, \hspace*{2cm} 
\label{Eq:vectorBSE}
\end{eqnarray}
with $p_{\hbox{\scriptsize out}}$ and $p_{\hbox{\scriptsize in}}$ the
outgoing and incoming quark momenta, respectively, and similarly for
$k_{\hbox{\scriptsize out}}$ and $k_{\hbox{\scriptsize in}}$, with
$p_{\hbox{\scriptsize out}}-p_{\hbox{\scriptsize
in}}=k_{\hbox{\scriptsize out}}-k_{\hbox{\scriptsize in}}=Q$.  The
kernel $K$ is the same kernel as used in the meson BSE, defined under
Eq.~(\ref{Eq:genBSE}).

Note that solutions of the {\em homogeneous} version of
Eq.~(\ref{Eq:vectorBSE}) define $q\bar{q}$ vector meson bound states
with masses \mbox{$M_{\rm V}^2=-Q^2$} at discrete timelike momenta
$Q^2$.  It follows that $\Gamma_\mu$ has poles at those locations and,
in the neighborhood of $Q^2 = -M_{\rm V}^2$, behaves
like~\cite{Maris:1999bh}
\begin{eqnarray}
 \Gamma_\mu(p_{\hbox{\scriptsize out}},p_{\hbox{\scriptsize in}}) &\sim& 
 \frac{\Gamma_\mu^{\rm V}(p_{\hbox{\scriptsize out}},p_{\hbox{\scriptsize in}}) \; 
   f_{\rm V} \, M_{\rm V}}{Q^2 + M_{\rm V}^2} \; ,
\end{eqnarray}
where $\Gamma_\mu^{\rm V}$ is the $q\bar{q}$ vector meson BSA, and
$f_{\rm V}$ the corresponding electroweak decay constant.  The fact
that the dressed $q\bar{q}\gamma$-vertex exhibits these vector meson
poles explains the success of naive vector-meson-dominance [VMD]
models; the effects of intermediate vector meson states on
electromagnetic processes can be unambiguously incorporated by using
the properly dressed $q\bar{q}\gamma$-vertex rather than the bare
vertex $\gamma_\mu$~\cite{Maris:1999bh}.

\section{Discussion of numerical results 
\label{Sec:Num}}
Numerical solutions for the quark propagator, vector meson BSAs and
dressed quark-gluon vertex can now be used in Eq.~(\ref{Eq:triangle})
to calculate the electromagnetic form factors.  We explicitly solve
the respective (in)homogeneous BSEs for the meson BSAs and for the
$q\bar{q}\gamma$-vertex in the corresponding momentum frame, thus
avoiding any need for interpolation or extrapolation of the numerical
solutions of the BSEs.  This does mean that we have to solve the meson
BSE for each value of $Q^2$, but the advantages of not having to
extrapolate the numerical BSAs outweighs the additional numerical
effort of repeatedly solving the meson BSE.  Further details are given
in the appendix.

\subsection{Rho and $K^\star$ form factors}
Results for the charge radius, 
\begin{eqnarray}
 \langle r^2 \rangle &=& -6 \; \frac{\partial G_{\rm E}(Q^2)}
                                 {\partial Q^2} \Bigg|_{Q^2 = 0} \;,
\end{eqnarray}
the magnetic moment $\mu = G_{\rm M}(0)$, and the magnetic quadrupole
moment ${\cal Q} = G_{\rm Q}(0)$, are presented in
Table~\ref{Table:resultsrhoKstar}. Experimental data on these 
static properties is absent, and so results from other 
theoretical models are included in the table.
\begin{table}[bt]
\caption{ Results for the $\rho$ and $K^\star$ meson charge radii
$\langle r_{\rm V}^2\rangle$ in fm$^2$, magnetic moments, and quadrupole
moments, compared to other calculations.  For comparison, we also
include the results for the pseudoscalar charge radii, calculated
within the same model, where available, as well as the experimental
pseudoscalar charge radii.
\label{Table:resultsrhoKstar} }
\renewcommand{\tabcolsep}{0.2pc} 
\renewcommand{\arraystretch}{1.2} 
\begin{tabular}{@{}l|c|ccc}
$\pi$, $\rho$ meson  & \multicolumn{1}{c|}{$r^2_\pi$} 
              & \multicolumn{1}{c}{$r^2_\rho$} 
              & \multicolumn{1}{c}{$\mu$}
              & \multicolumn{1}{c}{$\cal Q$}  \\ \hline
current DSE~\cite{Maris:1999bh,Maris:2006ea}
   & 0.44 & 0.54 & 2.01 & -0.41 \\
previous DSE~\cite{Burden:1995ve,Hawes:1998bz} 
   & 0.31 & 0.37 & 2.69 & -0.84 \\
Covariant QM~\cite{deMelo:1997hh}
   &      & 0.37 & 2.14 & -0.79 \\
Lightcone QM~\cite{Choi:1997iq,Choi:2004ww}  
   & 0.43 & 0.27 & 1.92 & -0.43 \\
experimental~\cite{PDG} & 0.452(11) \\ \hline
$u\bar{s}$ meson
              & $r^2_K$ & $r^2_{K^\star}$ & $\mu$ &$\cal Q$\\ \hline
current DSE calc.
              &  0.38   & 0.43 &  2.23 & -0.38  \\
previous DSE~\cite{Hawes:1998bz} 
              &  0.28   & 0.29 &  2.37 & -0.62  \\
experimental~\cite{PDG} & 0.314(35) \\ \hline
$d\bar{s}$ meson
              & $r^2_K$ & $r^2_{K^\star}$ & $\mu$ &$\cal Q$\\ \hline
current DSE calc.
              & -0.09   & -0.08 & -0.26 &  0.01 \\
previous DSE~\cite{Hawes:1998bz} 
              & -0.03   & -0.05 & -0.40 &  0.11 \\
experimental~\cite{PDG} & -0.077(10) \\ \hline
$c\bar{c}$ meson                
              & $r^2_{\eta_c}$ & $r^2_{J/\psi}$ & $\mu$ &$\cal Q$\\ \hline
current DSE~\cite{Maris:2006ea}
              & 0.048(4)& 0.052(3) &  2.13(4) & -0.28(1) \\ 
lattice~\cite{Dudek:2006ej}
              & 0.063(1)& 0.066(2) &  2.10(3) & -0.23(2) 
\end{tabular}\\[2pt]
\end{table}

Our calculations give a significantly larger value for the charge
radius of the $\rho$ and the charged $K^\star$ mesons compared to a
previous DSE calculation~\cite{Hawes:1998bz}, see
Table~\ref{Table:resultsrhoKstar}.  This is the case not only for the
vector mesons, but also for the pion and kaon charge radii.  This
difference can largely be attributed~\cite{Maris:1999bh} to the fact
that we employ a dressed quark-photon vertex that has poles in the
neighborhood of $Q^2 = -M_{\rm V}^2$.  Thus the effects from
intermediate vector meson states are unambiguously included in our
calculation.  On the other hand, Ref~\cite{Hawes:1998bz} uses the
Ball--Chiu Ansatz for the $q\bar{q}\gamma$-vertex.  Since our results
compare favorably with the experimental charge radii, we expect
our results for the vector radii to be more realistic than those of
Ref.~\cite{Hawes:1998bz}.

The quark model [QM] calculations of Refs.~\cite{deMelo:1997hh}
and \cite{Choi:2004ww} report a considerably smaller value for
$\langle r_\rho^2\rangle$.  Again, this is in part due to the fact
that neither of these calculations incorporate the effects of vector
meson poles in the quark-photon vertex~\cite{Maris:1999bh}.

Another significant difference between our results and those of
Ref.~\cite{Hawes:1998bz} is that the latter predicts a zero crossing
of $G_{\rm E}(Q^2)$ at about $Q^2 \approx 1.7~{\rm GeV}^2$ for the
$\rho$-meson, whereas we do not find any evidence for such a zero
crossing below \mbox{$Q^2 = 2.5~{\rm GeV}^2$}.  Extrapolating our
numerical results for $G_{\rm E}(Q^2)$ using a (2,3)-Pad\'e fit
suggests a zero crossing around $Q^2 \approx 3.8~{\rm GeV}^2$.  For
comparison, Ref.~\cite{deMelo:1997hh} predicts a zero crossing at $Q^2
\approx 2.9~{\rm GeV}^2$, and Ref.~\cite{Choi:2004ww} at $Q^2 \approx
5.5~{\rm GeV}^2$, whereas neither of the sum rule analyses 
\cite{Braguta:2004kx} and Ref.~\cite{Aliev:2004uj} predict a zero
crossing of $G_{\rm E}(Q^2)$ below $Q^2 = 5~{\rm GeV}^2$.  Based on
our calculations, and on the rather wide range of predictions from
other calculations, we conclude that it is unlikely for $G_{\rm
E}(Q^2)$ to have a zero crossing below \mbox{$Q^2 \approx 3~{\rm
GeV}^2$}.

The values we obtain for the magnetic and quadrupole moments of the
$\rho$-meson, $\mu$ and $\cal Q$, are very similar to those of
Ref.~\cite{Choi:2004ww}, and significantly smaller than those of
\cite{deMelo:1997hh} and Refs.~\cite{Hawes:1998bz}. However, the
$Q^2$ evolution of both $G_{\rm M}$ and $G_{\rm Q}$ is quite different
than that of Ref.~\cite{Choi:2004ww} (see
Table~\ref{Table:resultsQ2}), again most likely due to the fact that
VMD effects are not properly accounted for in Ref.~\cite{Choi:2004ww}.

\begin{table}[bt]
\caption{Our results for the $\rho$ meson electric, magnetic, and
quadrupole form factors $G_{\rm E,M,Q}(Q^2)$ at $Q^2=1~{\rm GeV}^2 $
and at $Q^2=2~{\rm GeV}^2 $, compared to previous
DSE~\cite{Hawes:1998bz}, light-cone~\cite{Choi:2004ww}, and sum rule
calculations~\cite{Aliev:2004uj}.
\label{Table:resultsQ2} }
\renewcommand{\tabcolsep}{0.6pc} 
\renewcommand{\arraystretch}{1.2} 
\begin{tabular}{@{}l|ccc|ccc}
 & \multicolumn{3}{c|}{$Q^2=1~{\rm GeV}^2 $}  
 & \multicolumn{3}{c}{$Q^2=2~{\rm GeV}^2 $}   \\ \hline
                         & \multicolumn{1}{c}{$G_{\rm E}$}
                         & \multicolumn{1}{c}{$G_{\rm M}$}
                         & \multicolumn{1}{c|}{$G_{\rm Q}$}
                         & \multicolumn{1}{c}{$G_{\rm E}$}
                         & \multicolumn{1}{c}{$G_{\rm M}$}
                         & \multicolumn{1}{c}{$G_{\rm Q}$} \\ \hline
current                  & 0.22 & 0.57 & -0.11 & 0.08 & 0.27 & -0.05 \\
Ref.~\cite{Hawes:1998bz} & 0.17 & 0.85 & -0.51 &-0.04 & 0.45 & -0.32 \\
Ref.~\cite{Choi:2004ww}  & 0.38 & 0.93 & -0.23 & 0.18 & 0.59 & -0.15 \\
Ref.~\cite{Aliev:2004uj} & 0.25 & 0.58 & -0.49 & 0.13 & 0.28 & -0.24 \\ 
\end{tabular}\\[2pt]
\end{table}
A recent sum rule analysis~\cite{Samsonov:2003hs} obtained $\mu = 2.0
\pm 0.3$ for the magnetic moment, which, given the large error
bars, is not accurate enough to discriminate between different
calculations.  The sum rule analysis of $\rho$ form factors at $Q^2 =
1~{\rm GeV}^2$ and $2~{\rm GeV}^2$~\cite{Aliev:2004uj} seems to
support our calculation, see Table~\ref{Table:resultsQ2}, at least for
$G_{\rm E}$ and $G_{\rm M}$.  Our results for the quadrupole form
factor however are almost a factor of five smaller than those of
Ref.~\cite{Aliev:2004uj}.  Clearly the quadrupole form factor is most
sensitive to the details of the dynamics.

\begin{figure}[tb]
\includegraphics[width=0.92\columnwidth]{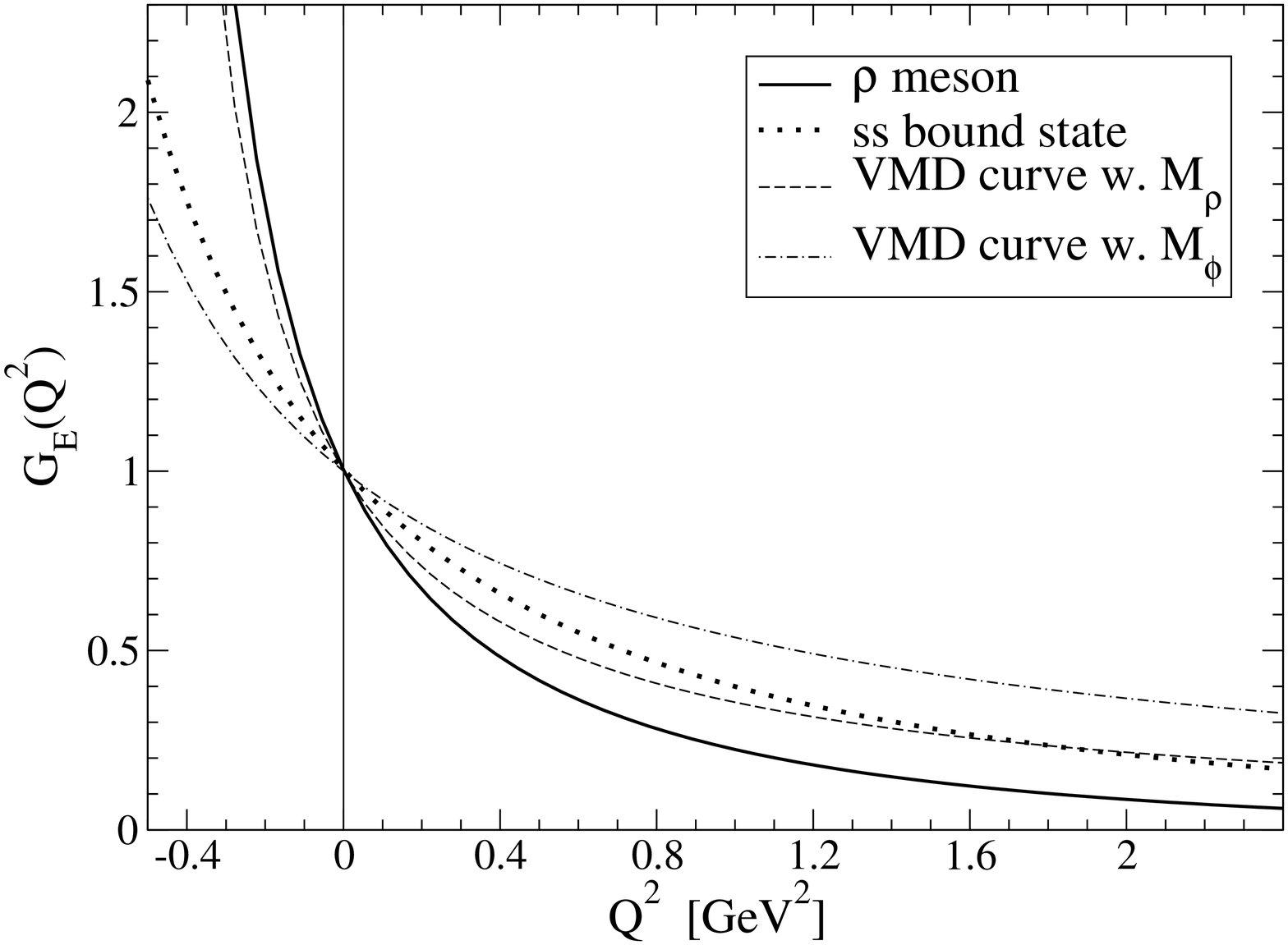}
\includegraphics[width=0.92\columnwidth]{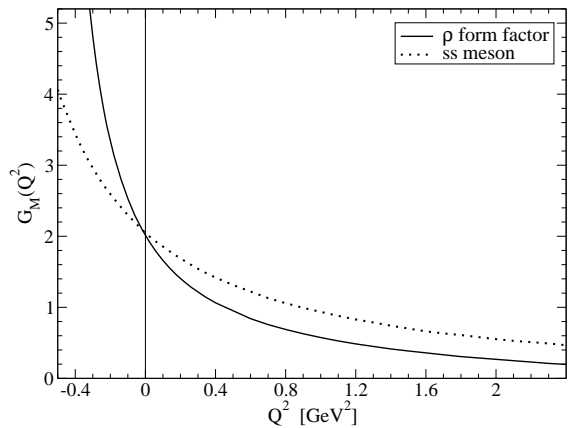}
\includegraphics[width=0.92\columnwidth]{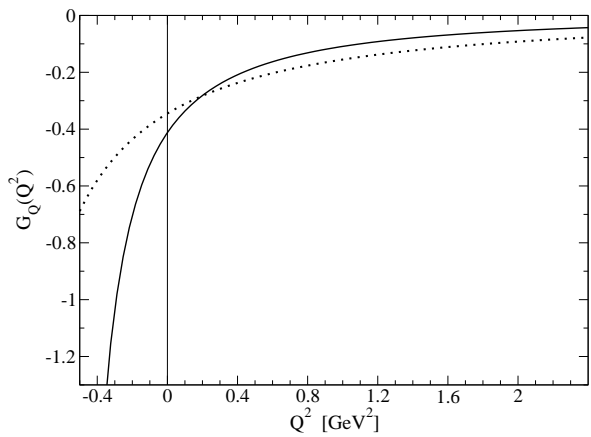}
\caption{
Numerical results for $G_{\rm E}(Q^2)$ (top), $G_{\rm M}(Q^2)$
(middle), and $G_{\rm Q}(Q^2)$ (bottom) for the $\rho$ meson and a
(fictitious) $s\bar{s}$ state.  For comparison, we also show the VMD
result (dashed and dot-dashed curves) for $G_{\rm E}(Q^2)$.
\label{Fig:rhoss}}
\end{figure}
\begin{figure}[tb]
\includegraphics[width=0.92\columnwidth]{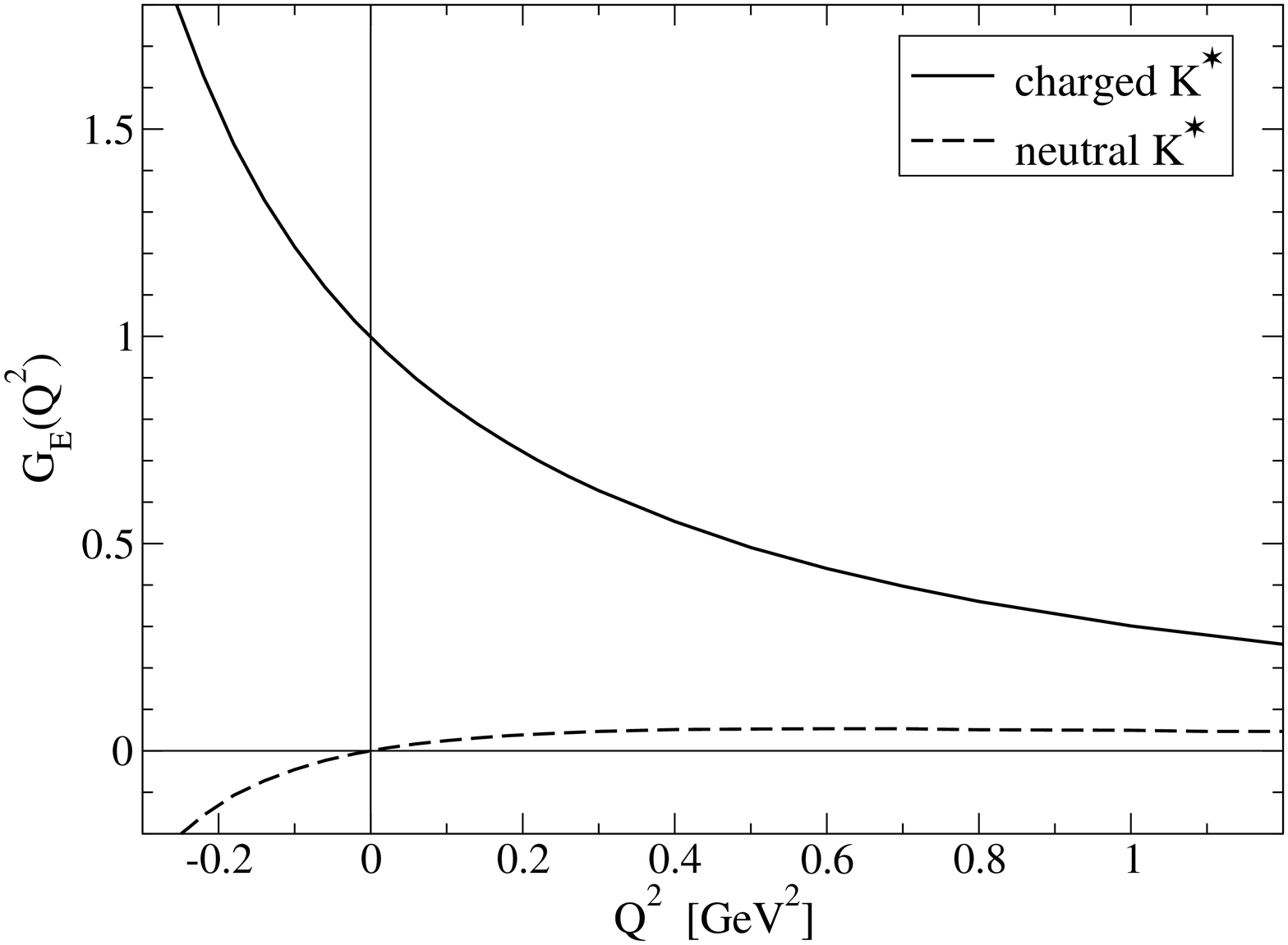}
\includegraphics[width=0.92\columnwidth]{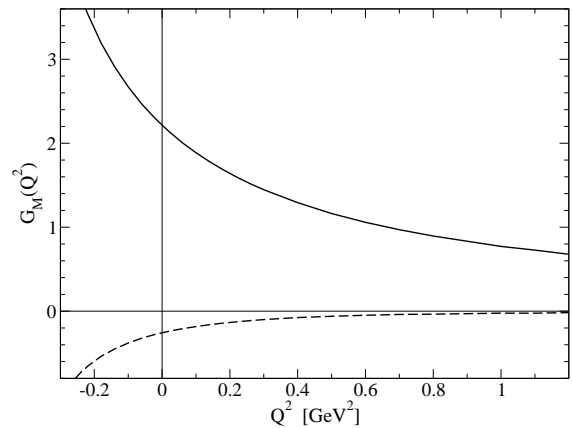}
\includegraphics[width=0.92\columnwidth]{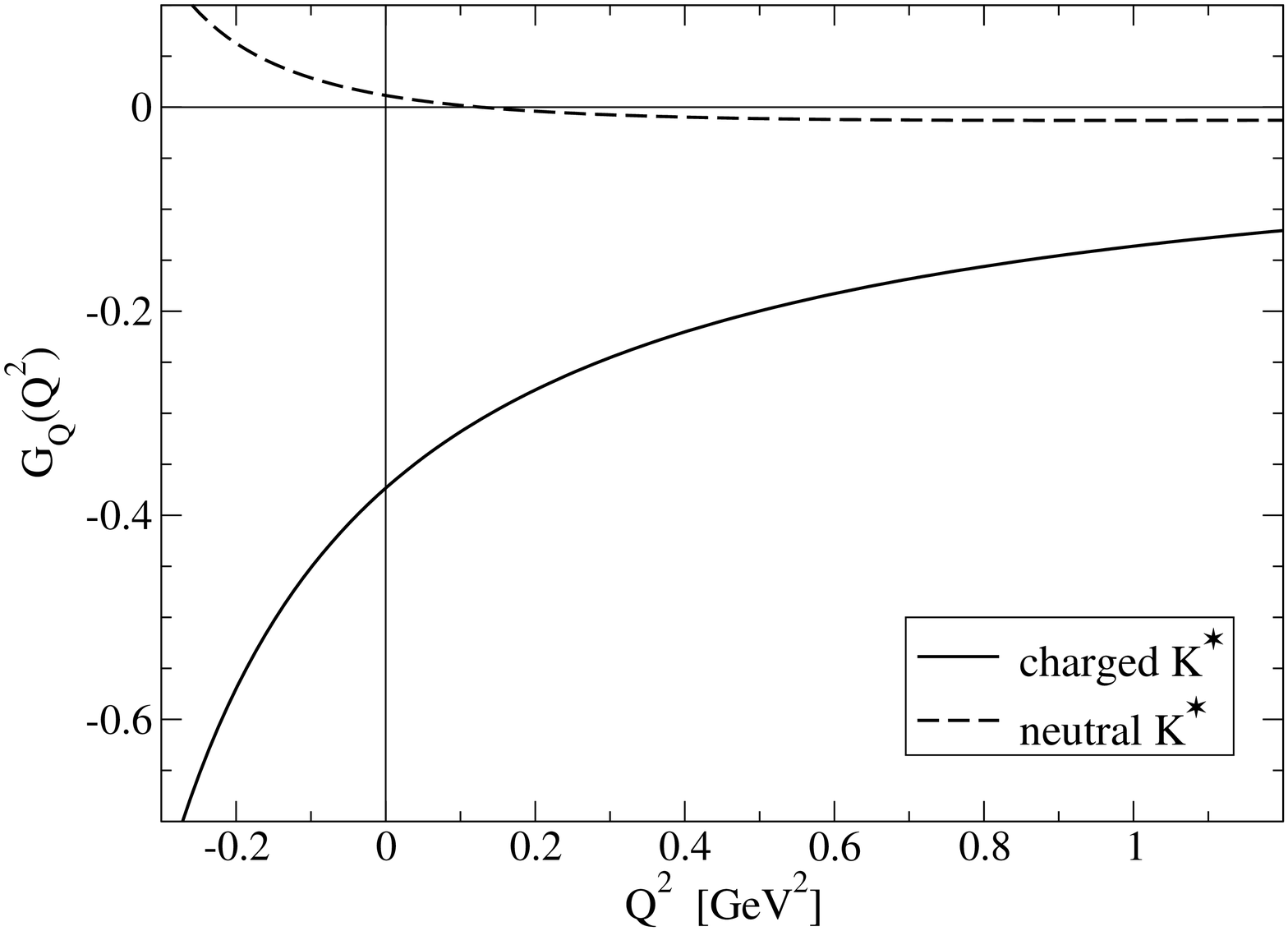}
\caption{ Numerical results for the neutral and charged $K^\star$ form
factors $G_{\rm E}(Q^2)$ (top), $G_{\rm M}(Q^2)$ (middle), and $G_{\rm
Q}(Q^2)$ (bottom).
\label{Fig:Kstar}}
\end{figure}
As we can see in Figs.~\ref{Fig:rhoss} and \ref{Fig:Kstar}, not only
$G_{\rm E}(Q^2)$, but also $G_{\rm M}(Q^2)$ and $G_{\rm Q}(Q^2)$ have
poles in the timelike region as one approaches $Q^2=-M_\rho^2$; in
addition, the $K^\star$ form factors also have poles at
\mbox{$Q^2=-M_\phi^2$}.  In Fig.~\ref{Fig:rhoss} we also show the
electromagnetic form factors of a fictitious $s\bar{s}$-like vector
meson with the photon coupled (with charge one) to only the $s$-quark,
but not to the $\bar{s}$-quark\footnote{The form factors of the
physical $\phi$ meson are trivially zero, as are the form factors of
the neutral $\rho$ meson.}.  The form factors of this fictitious
$s\bar{s}$-like vector meson have poles at \mbox{$Q^2=-M_\phi^2$}.
For comparison we also show a pure VMD model for the electric form
factor
\begin{eqnarray}
  G_{\rm E}(Q^2) &=&  \frac{M_{\rm V}^2}{M_{\rm V}^2 + Q^2} \;,
\end{eqnarray}
for both the $\rho$ and the $s\bar{s}$-like meson in the top panel of
Fig.~\ref{Fig:rhoss}.  Such a VMD model works remarkably well for the
pion electromagnetic form factor, at least up to spacelike $Q^2$
values of about 4 GeV$^2$.  Fig.~\ref{Fig:rhoss} shows that our form
factors of the vector mesons deviate significantly from a simple
VMD curve.  Only in the timelike region, near the actual vector meson
pole, does the VMD curve resemble our results.  In the spacelike
region, our results drop significantly faster than a VMD form factor.

Our results for the charged $K^\star$ meson form factors are
qualitatively similar to those for the $\rho$ meson.  The neutral
$K^\star$ form factors are most sensitive to details of the
calculation, since it depends on the difference between the up/down
quarks and the strange quarks.  Therefore we expect these form factors
to show more model dependence than the presented form factors of the
charged $K^\star$ and $\rho$ mesons, and this is indeed what we see in
Table~\ref{Table:resultsrhoKstar}, again in particular for the
quadrupole moment.

Finally, from Table~\ref{Table:resultsrhoKstar} it is interesting to
note that all studies, with the exception of the light-cone QM
calculation of Ref.~\cite{Choi:2004ww}, find the charge radii of
(charged) vector mesons to be larger than those of the corresponding
pseudoscalar mesons.  This trend was recently confirmed in lattice
calculations for light quarks~\cite{Lasscock:2006nh} and for
charmonium-like states~\cite{Dudek:2006ej}, with the photon coupled to
the quark only, not to the antiquark, of the charmonium state.  Also a
recent non-relativistic QM calculation of such charmonium-like states
gives a vector charge radius that is larger than the pseudoscalar
charge radius~\cite{Lakhina:2006vg}.  This means that the vector
states are broader than the corresponding pseudoscalar states,
assuming that the charge distribution is indicative of the physical
size of the bound state.  This agrees with the naive intuition that a
more tightly bound state is more compact than a heavier state with the
same constituents.

\subsection{Quark-mass dependence}
In addition to the electromagnetic form factors of the physical $\rho$
meson, we have also calculated the quark-mass dependence of these form
factors.  For simplicity we restrict ourselves to equal-mass mesons,
i.e. $q\bar{q}$ bound states, and for the electromagnetic properties
we couple the photon to only the quark $q$ (with charge one), not to
the antiquark $\bar{q}$.  We refer to these form factors as the
single-quark form factors; even though they are unphysical, they are
well-defined and allow for comparisons with other theoretical and
computational studies of the vector meson form factors.  

\begin{figure}[tb]
\includegraphics[width=0.96\columnwidth]{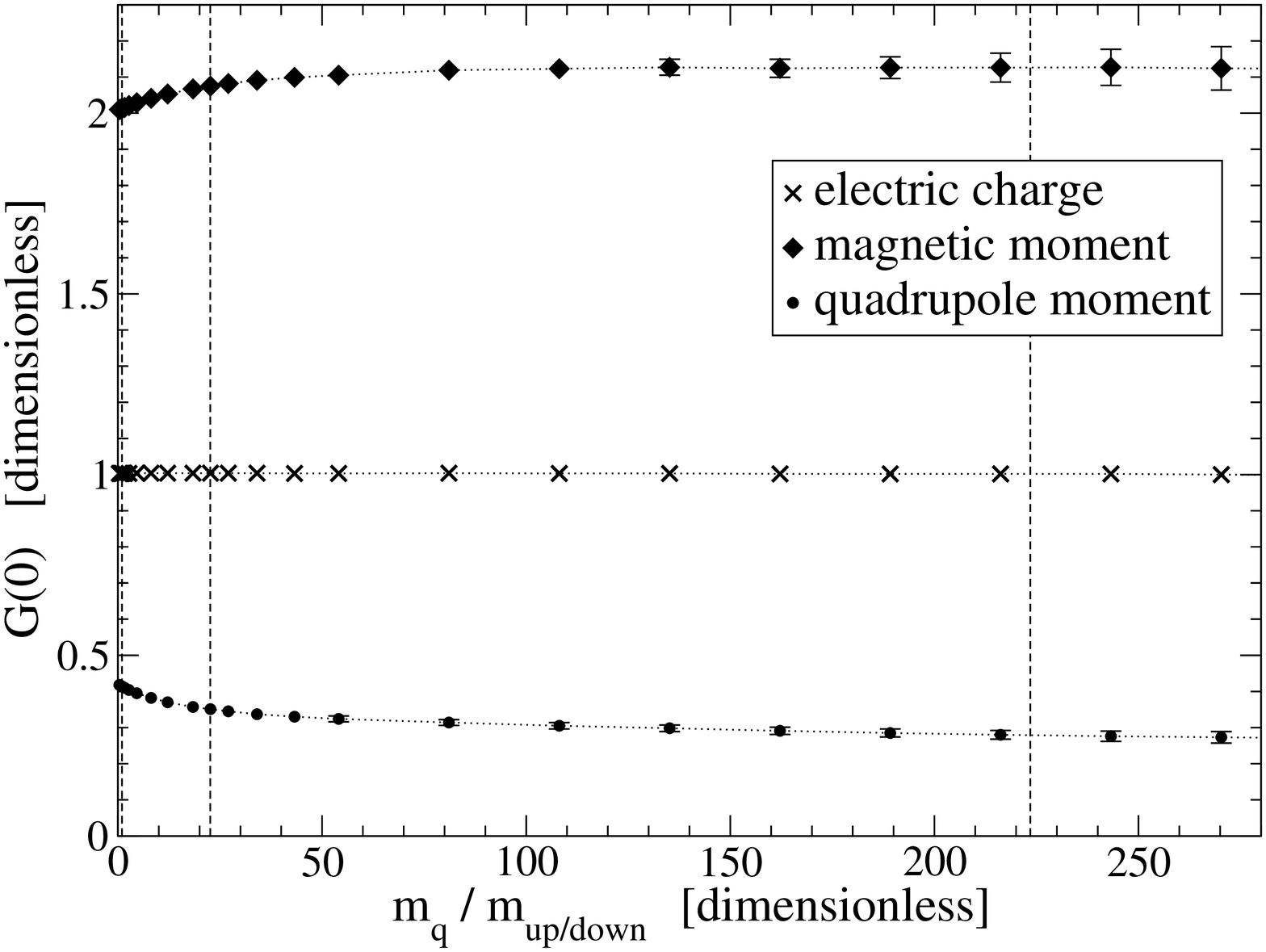}
\includegraphics[width=0.96\columnwidth]{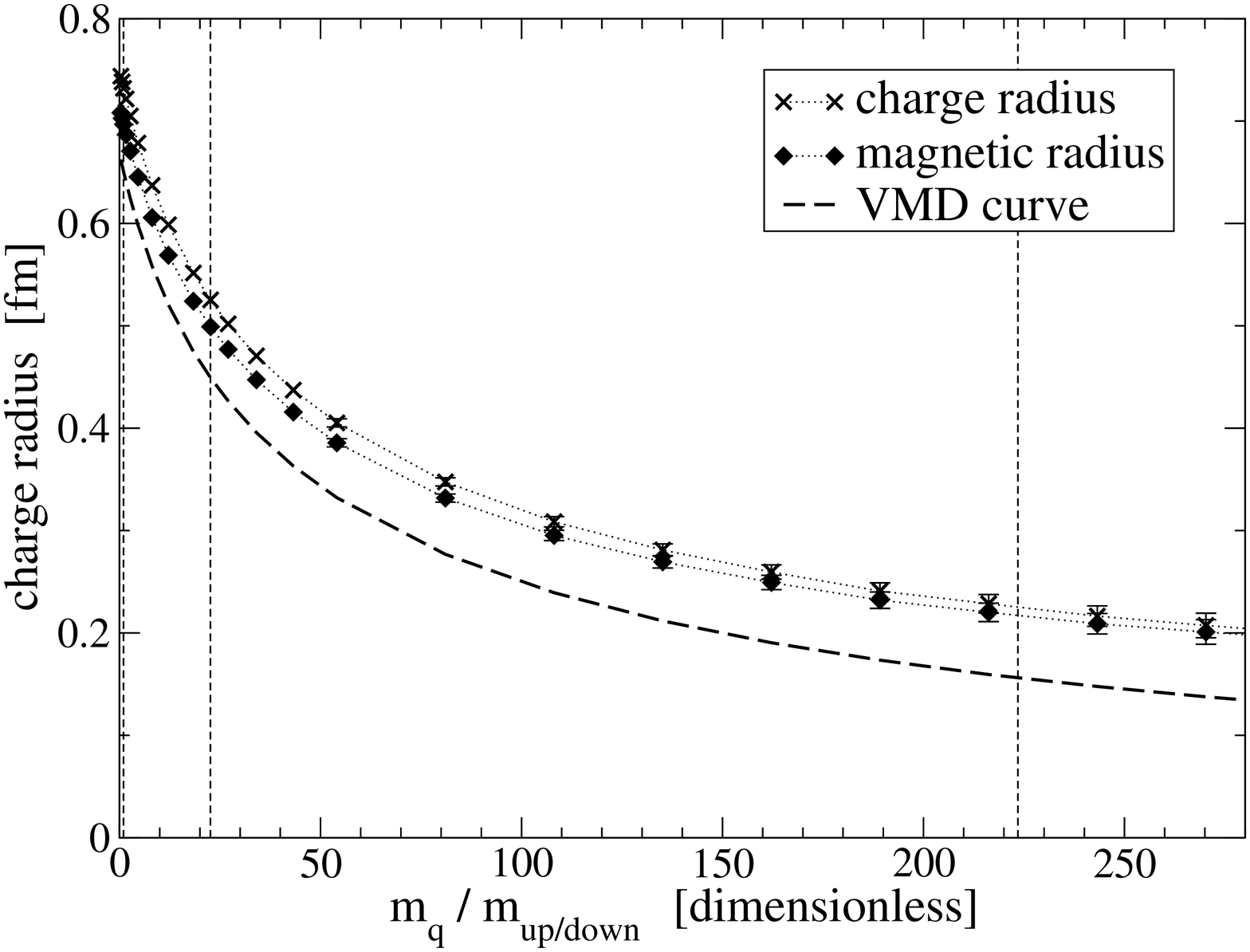}
\caption{Numerical results for the magnetic and quadrupole moments
(dimensionless, top) and for the charge radius (in fermi, bottom) as
function of the current quark mass.  Vertical dashed lines indicate
the up/down, strange, and charm quark masses.
\label{Fig:massdep}}
\end{figure}
The static electromagnetic properties are plotted in
Fig.~\ref{Fig:massdep} as function of the current quark mass,
normalized by the up/down current quark mass; for the strange quark we
have $m_s \approx 23~m_{u/d}$ and for the charm mass $m_c \approx
224~m_{u/d}$.  Our numerical errors grow with increasing meson mass
(i.e. increasing quark mass), mainly because the momentum $p^2$ of the
quark propagator in the Bethe--Salpeter integrals and in the triangle
diagram spans an increasingly large domain in the complex $p^2$-plane
with increasing meson mass.  The analytic continuation of the solution
of the quark DSE, Eq.~(\ref{Eq:quarkDSE}), from the spacelike
(Euclidean) axis to this complex momentum domain becomes numerically
cumbersome and inaccurate for large meson masses.  In our calculations
this is reflected by the fact that for the charm quark we have an
estimated 2\% to 4\% numerical uncertainty in the moments, and a
numerical uncertainty of 6\% to 8\% in the radii.

The magnetic moment $\mu$ turns out to be almost independent of the
current quark mass, Fig.~\ref{Fig:massdep}.  For $0< m_q < 4~m_s
\approx 100~m_{u/d}$ the magnetic moment increases with quark mass
from $\mu=2.01$ to $\mu=2.12$; above $4~m_s$ the numerical uncertainty
starts to increase and the magnetic moment is basically independent of
the quark mass within our numerical error bars of a few percent.

This mass dependence can naively be understood by considering a simple
constituent quark model, in which a vector meson is a bound state of
two quarks in an S-wave with the spins aligned.  In such a model one
expects the magnetic moment to be proportional to that of its
constituents, i.e. $\mu_{\rm V} \propto \mu_q$, where $\mu_q$ is the
magnetic moment of the constituent quark.  As long as the quark
magnetic moment is only weakly dependent on the current quark mass, so
is the magnetic moment of the bound state.

Since both the charge, $G_{\rm E}(0)$, and the magnetic moment,
$G_{\rm M}(0)$, are (almost) independent of the quark mass, one might
expect that also the quadrupole moment, $G_{\rm Q}(0)$, depends only
weakly on the quark mass.  However, our calculation shows that this is
not the case: The quadrupole moment ${\cal Q}$ decreases monotonically
with $m_q$, and is reduced by about 25\% at $m_q \approx 2~m_{s}
\approx 50~m_{u/d}$.  Above this quark mass, the quadrupole moment
continues to decrease with quark mass, but at a slower rate.  The mass
dependence of the quadrupole moment may not be that surprising on
realizing that our results, $-0.41 < {\cal Q} < -0.27$ depending on
quark mass, deviate significantly from the canonical value, ${\cal
Q}=-1$.  Furthermore, of all the static electromagnetic properties,
the quadrupole moment shows the largest model dependence, see
Table~\ref{Table:resultsrhoKstar}, indicating that the details of the
dynamics are important for this quantity.

The fact that we find a nontrivial quadrupole moment for all quark
masses, indicates that the naive picture of a vector meson as a
nonrelativistic bound state of two quarks in an S-wave is too simple.
Clearly, there is a significant amount of quark orbital momentum in
the vector mesons, which in our approach is incorporated in the meson
BSAs.  This is related to the fact that we use method that is
explicitly frame-independent: orbital angular momentum is not a
Poincar\'e invariant quantity.  

In the bottom panel of Fig.~\ref{Fig:massdep} we show the mass
dependence of the charge and magnetic radii (defined analogously) of
the vector mesons.  Just like the pion charge
radius~\cite{Maris:2005tt}, both the charge and the magnetic radius
decreases with increasing quark mass.  Our results for $\langle
r^2\rangle^{\frac{1}{2}}$ are qualitatively similar to a VMD curve,
and over the entire mass range the charge radius can in fact be
reasonably well described by a VMD curve with a constant shift of
about $0.07~{\rm fm}$.  Of course that means that the relative
deviation from a VMD model increases with increasing quark mass, again
just as in the case of the pion form factor.

Finally, we consider the single-quark transition form factors of the
$\eta_c$ and $J/\Psi$ mesons, at $m_c \approx 224 m_{u/d}$.  This
allows us to compare our results with the lattice simulations of
Ref.~\cite{Dudek:2006ej}.  Both the magnetic moment and the quadrupole
moment are in good agreement with the lattice data, see the bottom row
of Table~\ref{Table:resultsrhoKstar}.  Also the charge radii are in
reasonable agreement, with the vector state being slightly larger than
the pseudoscalar state.

In this paper we are interested in the quark-core contribution to
mesonic observables, and do not include pion loop effects.  However,
the form factors, and their quark-mass dependence, are sensitive to
pion loops, in particular at small quark masses.  Pion loops lead to
corrections to the physical $\rho$ and $\pi$ charge radii of about
10\% to 15\%~\cite{Alkofer:1993gu,Pichowsky:1999mu}, but their
contributions decrease rapidly with increasing quark mass.  Similarly,
we expect that at the physical up/down quark masses, the other form
factors, $G_{\rm M}$ and $G_{\rm Q}$ will also receive corrections of
the order of 10\% due to pion loops.  That means that the mass
dependence of the magnetic moment will be dominated by the pion loop
corrections, given the very weak quark-mass dependence we find here.
However, we expect these corrections to become negligible for masses
around $m_s$ and above.

\section{Conclusions
\label{Sec:Conc}}
We have calculated the electromagnetic form factors of the $\rho$
meson, and of both the charged and the neutral $K^\star$ mesons.  Our
method is explicitly Poincar\'e invariant, and we have demonstrated
explicitly that physical observables are frame-independent.  By
dressing the quark-photon vertex we guarantee electromagnetic current
conservations; furthermore, our dressed quark-photon vertex exhibits
poles in the timelike region, corresponding to intermediate vector
mesons.  Exactly the same method, using the same model for the
effective interaction~\cite{Maris:1999nt}, has been used quite
successfully to describe the pion electromagnetic form
factor~\cite{Maris:1999nt,Maris:2000sk} and a plethora of other light
meson observables within about 10\% to 15\%~\cite{Maris:2003vk}.  We
therefore expect our results to have a similar level of accuracy.

Compared to other calculations, we find a stronger $Q^2$-dependence,
mostly due to the fact that we have incorporated unambiguously VMD
effects in the quark-photon vertex, which have been neglected in
Refs.~\cite{Hawes:1998bz,Choi:2004ww,deMelo:1997hh}.  Our results
favor a magnetic moment close to two, in agreement with sum rule
analysis~\cite{Aliev:2004uj,Samsonov:2003hs}, and a quadrupole moment
${\cal Q} \approx -0.4$, which is similar to the result of
Ref.~\cite{Choi:2004ww}, but significantly smaller than suggested by
other calculations~\cite{Hawes:1998bz,deMelo:1997hh,Aliev:2004uj}.
Our results for the quadrupole moment indicate that there is a
significant amount of quark orbital angular momentum in the vector
mesons.

The magnetic moment is almost independent of the quark mass for mesons
of equal-mass constituents.  The quadrupole moment decreases with
increasing quark mass.  Also the shape of the form factors changes:
since the VMD pole in the timelike region shifts further away from
$Q^2=0$ with increasing quark mass, the form factors become less
steep, and the radii decrease with increasing quark mass.  Over the
entire quark mass range, from the chiral limit to the charm quark
mass, the charge radius can be reasonably described by a VMD curve
with a constant shift of about $0.07~{\rm fm}$.  A similar behavior of
the form factors, and of their quark-mass dependence, has been
obtained using the model of Ref.~\cite{Alkofer:2002bp} for the
effective quark-antiquark interaction.

Ideally experiments would guide us in order to discriminate between
different models and different calculation methods, but it is unlikely
that these form factors can be measured in the near future.  Lacking
reliable experimental input, it would be very useful to have
(quenched) lattice data for the vector meson form factors at light
quark masses.  Currently, the only accurate lattice data available are
at the charm mass~\cite{Dudek:2006ej}; these lattice data are in
reasonable agreement with our results at $m_c$.  Accurate lattice
simulations at light quark masses are needed in order to discriminate
between different model calculations.  A beginning has been made in
Ref.~\cite{Lasscock:2006nh}, but the error bars have to be reduced
significantly in order to make a detailed comparison meaningful.

\section{Acknowledgments}
We would like to thank Jo Dudek, Eric Swanson, Peter Tandy, and in
particular Craig Roberts for useful discussions and encouragement.
This work was supported by the US Department of Energy, contract
No.~DE-FG02-00ER41135, and the Department of Energy, Office of Nuclear
Physics, contract No.~DE-AC02-06CH11357. The work benefited from the
facilities of the NSF Terascale Computing System at the Pittsburgh
Supercomputing Center.

\appendix
\section{Numerical approach}
The form factors are calculated in impulse approximation
\begin{eqnarray}
\Lambda^{aa\bar{b}}_{\mu,\ \rho\sigma}(P,Q) &=&
        i\,N_c \int_k \!{\rm Tr}\big[ \Gamma^{a}_\mu(k_-,k_+)
	  \, \chi_\rho^{a\bar{b}}(k_+,k_P) 
\nonumber \\ && {} \times
        S^b(k_P)^{-1} \, \bar\chi_\sigma^{\bar{b}a}(q,k_-)\big] \;,
\label{App:triangle}
\end{eqnarray}
with \mbox{$k_P = k-P/2$} and \mbox{$k_\pm = k+P/2 \pm Q/2$}.
That means that we need numerical solutions of the rainbow-ladder BSE
for
\begin{itemize}
\item quark-photon vertex
\item incoming vector meson
\item outgoing vector meson
\end{itemize}
in addition to the solution of the quark DSE in rainbow truncation.
If we choose the momentum routing and integration variables carefully,
we can arrange the integration grids such that we do not need any
interpolation or extrapolation in the the final (triangle) loop
integral for the form factor.  That significantly reduces numerical
errors, in particular possible systematic errors introduced by the
extrapolation.

We use the momentum frame
\begin{eqnarray}
 P_\mu &=& (0, 0, 0,P) \;,
\\
 Q_\mu &=& (Q, 0, 0, 0) \;,
\end{eqnarray}
with $P^2 + \quarter Q^2 = -M^2$.  Depending on the value of $Q^2$,
either $P$ or $Q$ or both are imaginary.  In addition, we have the
integration momentum
\begin{eqnarray}
 k_\mu &=& k (\cos(\theta), \sin(\theta)\cos(\phi), 
          \sin(\theta)\sin(\phi),0) \;,
\end{eqnarray}
and corresponding measure
\begin{eqnarray}
 \int \frac{d^4k}{(2\pi)^4} &=&  \int_0^\infty \frac{k^3 \; dk}{(2\pi)^3}
    \int_0^\pi \sin^2(\theta) \; d\theta
    \int_0^\pi \sin(\phi) \; d\phi \;.
\nonumber \\
\end{eqnarray}
We use Eqs.~(\ref{Eq:HPT1})-(\ref{Eq:HPT3}) for the general structure
of the coupling of a photon to an on-shell vector meson, and perform
the traces analytically to obtain expressions for the form factors
$F_i(Q^2)$.  Subsequently, we use Eqs.~(\ref{Eq:HPGE})-(\ref{Eq:HPGQ})
to convert the functions $F_i$ to the more conventional electric,
magnetic, and quadrupole form factors $G_{\rm E}$, $G_{\rm M}$, and
$G_{\rm Q}$.

\subsection{Quark-photon vertex}
The $q\bar{q}\gamma$-vertex is the solution of an inhomogeneous BSE,
which in ladder truncation we can write as
\begin{eqnarray}
\Gamma_\mu(k_-,k_+;Q) &=&  Z_2 \, \gamma_\mu + {} 
 \frac{4}{3}\int\!\!\frac{d^4q}{(2\pi)^4}\; 
 4\pi\alpha\big((k-q)^2\big) \;  
\nonumber \\ && {} \times  D_{\rho\sigma}(k-q) \;
 \gamma_\rho \; \chi_\mu(q_-,q_+;Q) \; \gamma_\sigma \;,
\nonumber \\ 
\end{eqnarray}
where 
\begin{eqnarray}
\chi_\mu(q_-, q_+;Q) &=& S(q_-)\;\Gamma_\mu(q_-, q_+;Q)\;S(q_+) \;,
\end{eqnarray}
with \mbox{$k_\pm = k+P/2 \pm Q/2$}, and similarly for $q_\pm$.  Both
$k$ and $q$ are real Euclidean vectors, and the incoming and outgoing
quarks are always of the same flavor, so we can drop any flavor
indices.

In order to solve such a BSE, we decompose the function $\chi$ into
its Dirac components, and project out a set of coupled integral
equations for its scalar component functions $F_i$.  (Equivalently,
one could decompose $\Gamma$ into its Dirac components, and solve the
coupled integral equations for its components, but it turns out that
solving the BSE for $\chi$ is is roughly factor of three faster than
solving the BSE for $\Gamma$.)  We solve these coupled integral
equations by iteration, after discretizing the angular and radial
variables.

The most general form of the quark-photon vertex requires twelve Dirac
structures.  Four of these covariants represent the longitudinal
components which are completely specified by the Ward--Takahashi
identity in terms of the (inverse) quark propagator and they do not
contribute to elastic form factors.  The transverse part of the vertex
$\chi$ can be decomposed into eight components
\begin{eqnarray}
\lefteqn{ \chi_\mu(k_-, k_+;Q) \;= }
\nonumber \\ && {}
  \sum_{i=1}^8 T_i (k+\half P, Q) \; F_i(k^2, \theta, \phi) \;,
\end{eqnarray}
where the functions $F_i$ depend on {\em two} angles, because we solve
it in exactly the same frame and variables as it is used in the loop
integral for the form factor; alternatively, we could write it as a
function of $k^2$, $k\cdot P$, and $k\cdot Q$.

\subsection{Vector meson BSAs}
The incoming vector meson, with momentum $P - \half Q$ and flavor
labels $a\bar{b}$, is the solution of the homogeneous BSE
\begin{eqnarray}
\lefteqn{ S_a^{-1}(k_+)\;
         \chi^{a\bar{b}}_\rho(k_+,k_P;P-\half Q)\;
          S_b^{-1}(k_P) \;=\;}
\nonumber \\ && {}
 \frac{4}{3}\int\!\!\frac{d^4q}{(2\pi)^4}\;
 4\pi\alpha\big((k-q)^2\big) \; D_{\alpha\beta}(k-q) \;
\nonumber \\ && {} \times
 \gamma_\alpha \; \chi^{a\bar{b}}_\rho(q_+,q_P;P-\half Q) \; \gamma_\beta \;,
\end{eqnarray}
with \mbox{$k_+ = k + P/2 + Q/2$}, $k_P = k - P/2$, and similarly for
$q_+$ and $q_P$.  The vector meson is on its mass-shell: $P^2 +
\quarter Q^2 = -M^2$.

Again, we decompose the transverse vertex function $\chi$ into eight
components, but use a slightly different decomposition than for the
quark-photon vertex, because the momentum arguments are different.
For the vector meson BSAs we use
\begin{eqnarray}
\lefteqn{ \chi_\rho(k_+,k_P;P-\half Q) \;=} 
\nonumber \\ && {}
  \sum_{i=1}^8 T_i (k_P, P-\half Q) \; F_i(k^2, \theta, \phi) \;,
\end{eqnarray}
i.e. we use the momentum partitioning where the total incoming meson
momentum flows into the outgoing quark leg; or in other words, we use
the incoming quark momentum as the ``relative momentum'' in the
decomposition ($\eta = 1$), though not as the integration momentum.
And again, the functions $F_i$ depend on {\em two} angles.  As a check
on our numerics, we calculate the leptonic decay constant in this
frame as well, which gives us an indication of the numerical errors in
the BSAs.

We do {\em not} explicitly use any algebraic relation between $\chi$
and $\bar\chi$
\begin{eqnarray}
 \bar{\chi}(p, P) &=& 
  \Big[ C^{-1} \; \chi(-p, -P) \; C \Big]^{\hbox{\scriptsize transpose}} \;,
\end{eqnarray}
because the arguments of $\chi$ and $\bar\chi$ are different.
Instead, we simply solve both for $\chi$ and for $\bar\chi$ in the
appropriate frame.  That is, we calculate the BSA of the outgoing
vector meson as the solution of
\begin{eqnarray}
\lefteqn{ S_b^{-1}(k_P) \;
         \bar{\chi}^{\bar{b}a}_\sigma(k_P,k_-;P+\half Q)
          S_a^{-1}(k_-) \;=\;  }
\nonumber \\ && {}
 \frac{4}{3}\int\!\!\frac{d^4q}{(2\pi)^4}\;
 4\pi\alpha\big((k-q)^2\big) \; D_{\alpha\beta}(k-q) \;
\nonumber \\ && {} \times
 \gamma_\alpha \; \bar{\chi}^{\bar{b}a}_\beta(q_P,q_-;P+\half Q) \; \gamma_\beta \;,
\end{eqnarray}
again with \mbox{$k_- = k + P/2 - Q/2$}, $k_P = k - P/2$, and
similarly for $q_-$ and $q_P$.  We solve this equation for
$\bar{\chi}$ in basically the same manner as the BSE for $\chi$.  This
way we avoid the need for interpolation and extrapolation on the
vertex functions in the triangle diagram for the electromagnetic form
factors.


\end{document}